\documentclass[a4paper,11pt]{article}
\pdfoutput=1 

\usepackage{jcappub} 

\usepackage{xcolor}
\usepackage{hyperref}
\usepackage{amsmath}
\usepackage{multirow}
\usepackage{graphicx}
\usepackage{dcolumn}
\usepackage{bm}
\usepackage{amssymb}
\usepackage{latexsym}
\usepackage{booktabs}
\usepackage{amsmath}
\usepackage{enumerate}
\usepackage{url}
\usepackage{subfigure}
\usepackage{afterpage}
\usepackage{placeins}
\usepackage{float}

\newcommand{\ksm}{~{\rm km}~{\rm s}^{-1}~{\rm Mpc}^{-1}}

\begin{document}

\title{Ultra-low-frequency gravitational waves from individual supermassive black hole binaries as standard sirens}

\author[a,b]{Ling-Feng Wang,}
\author[a]{Yue Shao,}
\author[a]{Si-Ren Xiao,}
\author[a]{Jing-Fei Zhang}
\author[a,c,d,\ast]{and Xin Zhang}\note[$\ast$]{Corresponding author.}

\affiliation[a]{Liaoning Key Laboratory of Cosmology and Astrophysics, College of Sciences, Northeastern University, Shenyang 110819, China}
\affiliation[b]{School of Physics and Optoelectronic Engineering, Hainan University, Haikou 570228, China}
\affiliation[c]{National Frontiers Science Center for Industrial Intelligence and Systems Optimization, Northeastern University, Shenyang 110819, China}
\affiliation[d]{MOE Key Laboratory of Data Analytics and Optimization for Smart Industry, Northeastern University, Shenyang 110819, China}

\emailAdd{wanglf@hainanu.edu.cn}
\emailAdd{shaoyue@stumail.neu.edu.cn}
\emailAdd{xiaosiren@stumail.neu.edu.cn}
\emailAdd{jfzhang@mail.neu.edu.cn}
\emailAdd{zhangxin@mail.neu.edu.cn}

\abstract{
Ultra-low-frequency gravitational waves (GWs) generated by individual inspiraling supermassive black hole binaries (SMBHBs) at the centers of galaxies may be detected by pulsar timing arrays (PTAs) in the future. These GW signals, which encode absolute cosmic distances, can serve as bright and dark sirens, potentially evolving into a precise cosmological probe. Here, we show that a PTA in the era of the Square Kilometre Array, comprising 100 millisecond pulsars, could potentially detect about 25 bright sirens and 41 dark sirens over a 10-year observation period. The bright sirens, combined with cosmic microwave background data, offer capabilities comparable to current mainstream joint cosmological observations for measuring the equation of state of dark energy. The dark sirens could achieve a measurement precision of the Hubble constant close to that of current distance-ladder observations. Our results suggest that ultra-low-frequency GWs from individual SMBHBs are of great significance in investigating the nature of dark energy and determining the Hubble constant.
}
\maketitle
\section{Introduction} 
\label{sec:intro}
Gravitational waves (GWs) are ripples in the fabric of spacetime, produced when large masses accelerate.
The detection of GW150914 \cite{LIGOScientific:2016aoc}, the first GW event of binary black hole coalescence, has marked the beginning of the era of GW astronomy. The luminosity distances of GW sources, which are encoded in the amplitudes of GW waveforms, can be inferred from GW observations. This method is commonly referred to as ``standard sirens" \cite{Schutz:1986gp}. The standard sirens with electromagnetic (EM) counterparts can be used as ``bright sirens" to directly constrain cosmological parameters via the distance-redshift relation \cite{Schutz:1986gp,Dalal:2006qt,DES:2020nay}. For standard sirens without EM counterparts, researchers can use GW signals to identify their potential host galaxies in galaxy catalogs. A statistical analysis of these galaxies' redshifts, together with the GW signals, can also provide constraints on cosmological parameters. Such GW data are usually referred to as ``dark sirens" \cite{Schutz:1986gp,DelPozzo:2011vcw,Chen:2017rfc}.

Typical sources of standard sirens are compact binary coalescences, including stellar-mass compact binaries and supermassive black hole binaries (SMBHBs). Stellar-mass compact binaries, such as binary neutron stars (BNSs) and stellar-mass binary black holes (SBBHs), can be detected by ground-based GW detectors in the frequency band $\mathcal{O}(10)$ -- $\mathcal{O}(10^3)$ Hz. BNS coalescences are expected to have EM counterparts and have been experimentally confirmed by the GW170817 event \cite{LIGOScientific:2017vwq}, which is the only available bright siren to date, providing a $\sim14\%$ measurement for the Hubble constant $H_0$. SBBH coalescences are commonly thought to have no EM counterparts but they can serve as dark sirens. 47 such GW sources from the Third LIGO-Virgo-KAGRA Gravitational-Wave Transient Catalog provide a $\sim19\%$ measurement of the Hubble constant using the dark siren method \cite{LIGOScientific:2021aug}.
In the future, the third-generation ground-based GW detectors (the Einstein Telescope \cite{Punturo:2010zz} and the Cosmic Explorer \cite{LIGOScientific:2016wof}) enable researchers to acquire numerous standard sirens of stellar-mass compact binaries \cite{Chen:2020dyt}.

Low-frequency GWs emitted by SMBHBs with masses of $10^4$ -- $10^8$ $M_\odot$ can be detected in the millihertz (mHz) frequency band by the planned space-borne GW observatories, e.g., the Laser Interferometer Space Antenna \cite{LISA:2017pwj}, Taiji \cite{Hu:2017mde}, and TianQin \cite{Luo:2020bls}. These SMBHBs may produce EM emissions due to their surrounding gas-rich environments and external magnetic fields \cite{Okamoto:2005hi,Haiman:2017szj}, and therefore they are also expected to serve as bright sirens \cite{Tamanini:2016uin,Cai:2017yww,Wang:2019tto,Zhao:2019gyk,Wang:2021srv}. Recent studies show that such SMBHBs can also serve as dark sirens and provide precise measurements for $H_0$ \cite{Wang:2020dkc,Zhu:2021bpp}.

Ultra-low-frequency GWs emitted by SMBHBs with masses of $10^8$ -- $10^{10}$ $M_\odot$ are anticipated to be detected in the nanohertz (nHz) frequency band by the natural galactic-scale detector consisting of an array of millisecond pulsars (MSPs), commonly known as ``pulsar timing array" (PTA). When GWs pass between pulsars and the Earth, they alter the paths of the pulsar signals, consequently impacting the times of arrival (ToAs) of radio pulses. Nanohertz GWs from individual inspiraling SMBHBs could be detected by monitoring the spatially correlated fluctuations of ToAs induced by GWs. With the concept proposed decades ago, there are several major PTA projects, namely, the Parkes Pulsar Timing Array \cite{Hobbs_2013}, the European Pulsar Timing Array \cite{Kramer:2013kea}, the North American Nanohertz Observatory for Gravitational Waves \cite{McLaughlin:2013ira}, the Chinese Pulsar Timing Array \cite{Xu:2023wog}, the Indian Pulsar Timing Array \cite{Tarafdar:2022toa}, and the MeerKAT Pulsar Timing Array \cite{Miles:2022lkg}. These PTA projects have also been combined to form the International Pulsar Timing Array (IPTA) \cite{Hobbs:2009yy} aimed at significantly enhancing sensitivities. So far, most of the efforts have been devoted to detecting the stochastic gravitational wave background (SGWB).
Recently, several PTA projects have detected stochastic signals consistent with the Hellings-Downs correlations \cite{Hellings:1983fr}, providing suggestive evidence supporting the existence of ultra-low-frequency SGWB \cite{NANOGrav:2023gor,Reardon:2023gzh,EPTA:2023sfo,Xu:2023wog}. This gives researchers confidence that GWs from individual SMBHBs could also be detected in the future.
Although challenging, detecting individual SMBHBs will yield immense scientific returns. The capability of detecting individual SMBHBs using PTAs has been investigated in Refs.~\cite{Sesana:2008xk,Yardley:2010kv,Lee:2011et,Zhu:2015tua,Wang:2016tfy,Xiao:2024nmi}. With the participation of more advanced radio telescopes such as the Five-hundred-meter Aperture Spherical Telescope (FAST) \cite{Nan:2011um} in China and the planned Square Kilometre Array
(SKA) \cite{Lazio:2013mea}, there is a great possibility that GWs produced by individual SMBHBs (other than SGWBs) could be detected by SKA-era PTAs \cite{Smits:2008cf}.

Recently, it was proposed in Ref.~\cite{Yan:2019sbx} that inspiraling SMBHBs to be detected by PTAs could also serve as bright sirens. The luminosity distances of currently available SMBHB candidates detected by EM observations \cite{Valtonen:2008tx} with known redshifts may be measured by the PTA GW observations. Subsequently, the distance-redshift relation can be used to constrain cosmological parameters.
In Ref.~\cite{Yan:2019sbx}, a preliminary study on constraining dark energy parameters was performed, in which only the equation-of-state (EoS) parameters of dark energy were allowed to vary, while all other cosmological parameters were fixed. Obviously, such a treatment cannot reveal how well the PTA nHz GW observations could constrain cosmological parameters.
Actually, the most prominent advantage of GW bright sirens in cosmological parameter estimations is that they can break the degeneracies between cosmological parameters \cite{Zhang:2019ylr,Zhang:2019loq,Jin:2020hmc,Wang:2021srv}. The capabilities of the bright sirens from ground-based detectors and space-borne observatories in breaking the parameter degeneracies have been widely discussed (see Ref.~\cite{Bian:2021ini} for a recent review), but the relevant studies on the standard sirens from PTA observations are still absent. Here, the first question to be answered is what role ultra-low-frequency GW bright sirens can play in breaking the degeneracies between cosmological parameters.

Although the SMBHB bright sirens from the PTA observations are considered valuable for measuring cosmological parameters, they also have limitations. This is because the SMBHB candidates with known redshifts may not actually be SMBHBs and the detected SMBHBs may not belong to these candidates. Therefore, it is important to find a way to measure cosmological parameters when the SMBHB bright sirens are not available. Here we propose for the first time that SMBHBs detected by PTAs may serve as dark sirens. Dark sirens require suitable galaxy catalogs to identify potential host galaxies of SMBHBs. Since the redshifts of the existing galaxy catalogs are relatively low, only SMBHBs in the local Universe might be used as dark sirens. Along this line, the second question we wish to answer is whether SMBHBs in the local Universe can be utilized as ultra-low-frequency GW dark sirens to precisely measure cosmological parameters.

In this work, we analyze the capability of SKA-era PTAs to detect the known SMBHB candidates and mock local-Universe SMBHBs by simulating the timing residuals of pulsar signals. Subsequently, we comprehensively analyze the role of both SMBHB bright and dark sirens in cosmological parameter estimations for the first time. The system of units in which $G=c=1$ is adopted in this paper.


\section{Detection of individual SMBHBs}
\label{sub:detection}

\begin{figure}
\centering
\includegraphics[angle=0, width=10cm]{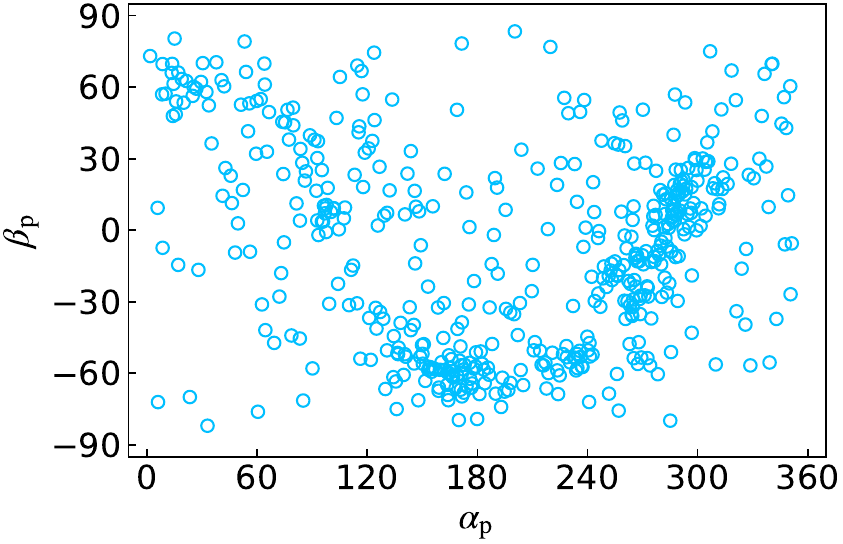}
\caption{\label{fig_MSP}
  Positions of the selected MSPs in the sky.
  We select 500 pulsars located within 3 kpc from the Earth from the ATNF pulsar catalog \cite{Manchester:2004bp}.}
\end{figure}

GW signals are detected in the timing residuals of MSPs by subtracting model-predicted ToAs from the observed ToA data. The timing residuals induced by a single GW source, measured at time $t$ on the Earth, can be expressed as
\begin{align}\label{s-definition}
s(t,\hat{\Omega}_{\rm s},\hat{\Omega}_{\rm p})=F_{+}(\hat{\Omega}_{\rm s},\hat{\Omega}_{\rm p})\Delta A_+(t)+F_{\times}(\hat{\Omega}_{\rm s},\hat{\Omega}_{\rm p})\Delta A_{\times}(t),
\end{align}
where $F_{+,\times}(\hat{\Omega}_{\rm s},\hat{\Omega}_{\rm p})$ represent geometric factors, equivalent to the antenna pattern functions of laser interferometric GW detections \cite{Yan:2019sbx},
\begin{align}
F_{+}(\hat{\Omega}_{\rm s},\hat{\Omega}_{\rm p})=\frac{1}{4(1-\cos\theta)}\left\{(1+\sin^2\beta_{\rm s})\cos^2\beta_{\rm p}\cos[2(\alpha_{\rm s}-\alpha_{\rm p})]\right.\nonumber\\
 \left. -\sin2\beta_{\rm s}\sin2\beta_{\rm p}\cos(\alpha_{\rm s}-\alpha_{\rm p})
+\cos^2\beta_{\rm s}(2-3\cos^2\beta_{\rm p})\right\},
\end{align}
\begin{align}
F_{\times}(\hat{\Omega}_{\rm s},\hat{\Omega}_{\rm p})=\frac{1}{2(1-\cos\theta)}\left\{\cos\beta_{\rm s}\sin2\beta_{\rm p}\sin(\alpha_{\rm s}-\alpha_{\rm p}) \right.
\left. -\sin\beta_{\rm s}\cos^2\beta_{\rm p}\sin[2(\alpha_{\rm s}-\alpha_{\rm p})]\right\}.
\end{align}
Here, $\hat{\Omega}_{\rm s}$ and $\hat{\Omega}_{\rm p}$ are the unit vectors pointing from the GW source and the pulsar to the observer, respectively. These vectors are determined by the sky positions of the GW source ($\alpha_{\rm s}$, $\beta_{\rm s}$) and the pulsar ($\alpha_{\rm p}$, $\beta_{\rm p}$).
$\theta$ represents the angular separation between the GW source and the pulsar,

\begin{align}
\cos\theta=\cos\beta_{\rm s}\cos\beta_{\rm p}\cos(\alpha_{\rm s}-\alpha_{\rm p})+\sin\beta_{\rm s}\sin\beta_{\rm p}.
\end{align}
Figure~\ref{fig_MSP} displays the sky positions of the 500 selected MSPs used in this study. $\Delta A_{+,\times}(t)=A_{+,\times}(t)-A_{+,\times}(t_{\rm p})$ represents the disparity between the Earth term $A_{+,\times}(t)$ and the pulsar term $A_{+,\times}(t_{\rm p})$, with $t_{\rm p}$ the time at which GW passes the MSP \cite{Ellis:2012zv},

\begin{align}
t_{\rm p} = t-d_{\rm p}(1-\cos\theta)/c,
\label{tpt}
\end{align}
where $d_{\rm p}$ represents the pulsar distance.

The specific forms of the Earth term and the pulsar term depend on the types of GW sources.
SMBHBs in circular orbits are expected to generate pulsar timing signals with the following forms,
\begin{align}
A_{+}(t)=&\frac{h(t)}{2\pi f(t)}\left\{(1+\cos^2\iota)\cos2\psi\sin[\phi(t)\right.
\left.+\phi_0]+2\cos\iota\sin2\psi\cos[\phi(t)+\phi_0]\right \}, \label{A+} \\
A_{\times}(t)=&\frac{h(t)}{2\pi f(t)}\left\{(1+\cos^2\iota)\sin2\psi\sin[\phi(t) \right.
\left. +\phi_0]-2\cos\iota\cos2\psi\cos[\phi(t)+\phi_0]\right \}.
\label{Across}
\end{align}
Here, $\iota$ is the inclination angle of the binary orbit, $\psi$ is the GW polarization angle, and $\phi_0$ is the phase constant.
We assume that SMBHBs inspiral in circular orbits, and then GW strain amplitude $h(t)$ can be expressed as
\begin{align}
h(t)=2\frac{(G\mathcal{M}_{\rm c})^{5/3}}{c^4}\frac{[\pi f(t)]^{2/3}}{d_{\rm L}}.
\end{align}
Here, $\mathcal{M}_{\rm c}=M_{\rm c}(1+z)$ represents the redshifted chirp mass, where $M_{\rm c}$ is the chirp mass defined as $M_{\rm c}=\eta^{3/5}M$. $M=m_{1}+m_{2}$ represents the total mass
of the binary system comprising component masses $m_{1}$ and $m_{2}$, and $\eta=m_{1}m_{2}/(m_{1}+m_{2})^{2}$ denotes the symmetric mass ratio. $d_{\rm L}$ represents the luminosity distance of the GW source.
The GW frequency $f(t)$ is given by
\begin{align}\label{ft}
f(t)=\left[f_0^{-8/3}-\frac{256}{5}\pi^{8/3}\left(\frac{G\mathcal{M}_{\rm c}}{c^3}\right)^{5/3}t\right]^{-3/8},
\end{align}
where $f_0=2f_{\rm orb}$ represents the GW frequency at the time of the first observation. Here $f_{\rm orb}=(2\pi {T_{\rm obs}})^{-1}$ is the orbit frequency, where ${T_{\rm obs}}$ stands for the orbital periods of SMBHBs. When simulating the GW signals of bright sirens, we calculate $f_0$ using the orbital periods of the 154 SMBHB candidates \cite{Graham:2015gma,Graham:2015tba,Charisi:2016fqw,Yan:2015mya,Li:2016hcm,Valtonen:2008tx,Zheng:2015dij,Li:2017eqf}. When simulating the GW signals of dark sirens, we calculate $f_0$ using eq.~(\ref{f0}).

The signal-to-noise ratio (SNR) of the GW signal detected by a PTA is determined by
\begin{align}\label{rho}
\rho^2=\sum_{i=1}^{N_{\rm p}}\sum_{n=1}^{N_t}\left[\frac{s_i(t_n)}{\sigma_{t,i}}\right]^2.
\end{align}
$N_{\rm p}$ represents the number of available MSPs in the SKA-era PTAs and is still uncertain \cite{Feng:2020nyw}. Currently, the second data released by IPTA contains 65 available MSPs \cite{Antoniadis:2022pcn}. Based on this number, we assume that the available MSPs in the SKA era could increase several times. Therefore, we select 100, 200, and 500 MSPs within 3 kpc from the Earth, obtained from the Australia Telescope National Facility (ATNF) pulsar catalog \cite{Manchester:2004bp}, to construct SKA-era PTAs. $N_t$ is the number of time samples determined by the cadence of monitoring the pulses from MSPs. Here, we set the cadence to two weeks and assume that the observation span is 10 years \cite{Yan:2019sbx}. $s_i(t_n)$ is the timing residual induced by the GW signal in the $i$-th MSP at time $t_n$ [see eq.~(\ref{s-definition})].
$\sigma_{t,i}$ is the root mean square (rms) of the timing residual of the $i$-th MSP. It comprises red noise and white noise, reflecting the stability of the pulsar and the quality of the ToA data.

The white noise mainly includes jitter noise and radiometer noise. The jitter noise would dominate for most bright pulsars and the total white noise is around 10 -- 50 ns \cite{Porayko:2018sfa}. Considering that FAST and SKA could reduce the noise, we anticipate that $\sigma_t$ could reach $\sim$ 20 ns for SKA-era PTAs. We consider two cases of $\sigma_t=20$ ns and $\sigma_t=100$ ns for comparison.
Here we assume that the GW spectrum induced by SGWB can be accurately measured in the upcoming years and the GW signals from individual SMBHBs can be distinguished from SGWB \cite{2019NatAs...3....8M}. Therefore, we do not take SGWB into account in this work.
Since the GW strain induced by an individual source in the frequency domain appears as a single peak on the PTA-detection time scale, which is fundamentally different from frequency-dependent SGWB \cite{Lee:2011et}, the red noise primarily impacts the detection of SGWB while it has a lesser impact on the detection of individual sources, especially at relatively high frequencies. Therefore, we ignore the influence of the red noise when obtaining our main results. To make our analysis more robust, we also conducted additional analyzes by considering red noise (see Section~\ref{sub:bright}).

The Fisher information matrix is used to estimate the parameters of GW sources.
For a PTA comprising $N_{\rm p}$ independent MSPs, the Fisher matrix $\boldsymbol{\mathcal{F}}$ is expressed as
\begin{align}\label{fisher}
{\mathcal{F}}_{ab}=\sum_{i=1}^{N_{\rm p}}\sum_{n=1}^{N_t}\frac{\partial{s_i(t_n)}}{\sigma_{t,i}\partial{\theta_a}}\frac{\partial{s_i(t_n)}}{\sigma_{t,i}\partial{\theta_b}},
\end{align}
where $\boldsymbol{\theta}$ denotes the free parameters to be estimated.
The instrumental error of the parameter $\theta_a$ is estimated as $\Delta \theta_a=\sqrt{(\mathcal{F}^{-1})_{aa}}$.
{Typically, $8+N_{\rm p}$ parameters are taken into account in the Fisher matrix, including 8 parameters describing the GW source (i.e., $M_{\rm c}$, $\alpha_{\rm s}$, $\beta_{\rm s}$, $\iota$, $\psi$, $\phi_0$, $f_0$, $d_{\rm L}$) and $N_{\rm p}$ parameters describing the pulsar distances. According to our calculations, there is a strong degeneracy between the parameters $d_{\rm L}$ and $M_{\rm c}$, unless the parameter $d_{\rm p}$ is fixed. This is because fixing $d_{\rm p}$ is equivalent to precisely knowing the time difference between the pulsar term and the Earth term [see eq.~(\ref{tpt})]. Although the variation of the GW frequency of an inspiraling SMBHB is minuscule over a 10-year observational period, the time difference between the pulsar term and the Earth term is not negligible. Therefore, the frequency difference between the pulsar term and the Earth term is actually significant. With this, $\mathcal M_{\rm c}$ can be inferred from the derivative of frequency with respect to time [see eq.~(\ref{ft})], thereby breaking the degeneracy between $\mathcal M_{\rm c}$ and $d_{\rm L}$. Here, we assume that the pulsar distance could be precisely measured in the SKA era. Therefore, we fix $d_{\rm p}$ and consider only the GW source parameters in the Fisher matrix.} The inclination angle $\iota$ is randomly chosen in [0, $\pi$]. The polarization angle $\psi$ and the initial phase $\phi_0$ of SMBHBs are randomly chosen in [0, $2\pi$]. 
In addition to the instrumental error ($\Delta d_{\mathrm{L}}^{\rm inst}$) estimated by the Fisher matrix, the total error of $d_{\rm L}$ should also include the weak lensing error ($\Delta d_{\mathrm{L}}^{\rm lens}$) \cite{Hirata:2010ba},
\begin{equation}
\Delta d_{\mathrm{L}}=
\sqrt{\left(\Delta d_{\mathrm{L}}^{\rm inst}\right)^2+
\left(\Delta d_{\mathrm{L}}^{\rm lens}\right)^2},
\end{equation}
with
\begin{equation}
\Delta d_{\mathrm{L}}^{\rm lens}(z)=d_{\mathrm{L}}(z) \times 0.066\left(\frac{1-(1+z)^{-0.25}}{0.25}\right)^{1.8}.
\end{equation}
 
{To evaluate the numerical stability of the Fisher matrix, we compute its 2-norm condition number, which is given by
\begin{equation}
\kappa(\boldsymbol{\mathcal{F}}) = \|\boldsymbol{\mathcal{F}}\|_2 \times \|\boldsymbol{\mathcal{F}}^{-1}\|_2 = \frac{\sigma_{\max}(\boldsymbol{\mathcal{F}})}{\sigma_{\min}(\boldsymbol{\mathcal{F}})},
\end{equation}
where \( \sigma_{\max}(\boldsymbol{\mathcal{F}}) \) and \( \sigma_{\min}(\boldsymbol{\mathcal{F}}) \) are the largest and smallest singular values of the Fisher matrix \( \boldsymbol{\mathcal{F}} \), respectively.}
A large condition number of the Fisher matrix will lead to the numerical instability of the matrix inversion. Through examinations, we find that many matrices have condition numbers much greater than 1 and the matrices approach singularity, which causes the matrix and its inverse not to multiply into a diagonal matrix. To address this issue, we normalize the matrices by using the method displayed in Ref.~\cite{Dupletsa:2022scg}, which significantly reduce the condition numbers of most matrices. Although some matrices still have large condition numbers due to the parameter degeneracy, our tests show that they are not singular. Moreover, we also calculate the constraint precision of $H_0$ after the matrix normalizations, and find that the constraint results do not change significantly.


\section{Bayesian Analysis}
In the Bayesian method, the posterior distribution of $H_0$ can be expressed as
\begin{equation}
    \begin{aligned}    p(H_0|\mathcal{D}_{\rm{GW}},\mathcal{D}_{\rm{EM}})\propto p(\mathcal{D}_{\rm{GW}},\mathcal{D}_{\rm{EM}}|H_0)p(H_0),
    \end{aligned}
\end{equation}
where $\mathcal{D}_{\rm GW}$ and $\mathcal{D}_{\rm EM}$ represent the GW and EM data, respectively. $p(H_0)$ represents the prior probability of $H_0$, assumed to be uniformly distributed in the interval [50, 80] $\ksm$.
For a single GW event, the likelihood term, $p(\mathcal{D}_{\rm{GW}},\mathcal{D}_{\rm{EM}}|H_0)$, can be written as
\begin{align}\label{con:likelihood}
p(\mathcal{D}_{\rm GW},\mathcal{D}_{\rm EM}|H_0)=\frac{\int p(\mathcal{D}_{\rm GW}|{d_{\rm L}}(z,H_0),\alpha, \beta)p(\mathcal{D}_{\rm EM}|z,\alpha, \beta)p_0(z,\alpha, \beta) {\rm d}z {\rm d}\alpha {\rm d}\beta}{\gamma (H_0)}.
\end{align}

$p(\mathcal{D}_{\rm GW}|d_{\rm L}(z,H_0),\alpha, \beta)$ in eq.~(\ref{con:likelihood}) represents the likelihood of the GW data, expressed as
\begin{align}
p(\mathcal{D}_{\rm GW}|{d_{\rm L}}(z,H_0),\alpha, \beta) \propto e^{-\chi^2 /2},
\end{align}
with $\chi^2 = (\boldsymbol{x}-\boldsymbol{x}_{\rm gw})^\mathrm{T} \boldsymbol{C}^{-1} (\boldsymbol{x}-\boldsymbol{x}_{\rm gw})$.
Here $\boldsymbol{x}=(d_{\rm L}(z,H_0),\alpha,\beta)$ represents an arbitrary three-dimensional (3D) position in the sky. $\boldsymbol{x}_{\rm gw}=(d_{\rm L,s}, \alpha_{\rm s}, \beta_{\rm s}$) represents the 3D position of the GW source, specifically, the 3D position of the true host galaxy in our simulation. Note that the actually measured values are in the error range of $\boldsymbol{x}_{\rm gw}$. We randomly select $\boldsymbol{x}_{\rm gw}$ from within its error range based on a Gaussian distribution.
We calculate $d_{\rm L,s}$ with the galaxies' redshifts ($z_s$) by assuming the $\Lambda$CDM model, where $\Omega_{\rm m}$ and $H_0$ are set to the $\emph{Planck}$ 2018 results.
$\boldsymbol{C}$ is the $3\times3$ covariance matrix relevant only to ($d_{\rm L}$, $\alpha$, $\beta$), obtained from the Fisher matrix.
We use $\chi^2=11.34$ (corresponding to 99\% confidence level) to determine the boundary of GW source's localization volume. For dark sirens, if the position of a galaxy satisfies $\chi^2<11.34$, we consider this galaxy to be within the localization volume and regard it as a potential host galaxy of the GW source.

$p(\mathcal{D}_{\rm EM}|z,\alpha, \beta)$ in eq.~(\ref{con:likelihood}) is the likelihood of the EM data. {For bright sirens, we assume that their host galaxies could be uniquely identified through the EM counterparts. In this case, the likelihood term \( p(\mathcal{D}_{\rm EM} | z, \alpha, \beta) \) is expressed as
\begin{equation}
p(\mathcal{D}_{\rm EM} | z, \alpha, \beta) = \mathcal{N}(z_s, \sigma_{z, s}) \delta(\alpha - \alpha_s) \delta(\beta - \beta_s), 
\end{equation}
where \( z_s, \alpha_s, \beta_s \) represent the 3D position of the host galaxy of the GW source, derived from the EM counterpart. 
\( \mathcal{N}(z_s, \sigma_{z, s}) \) is a Gaussian distribution with mean \( z_s \) and standard deviation \( \sigma_{z_s} = (1 + z_s) \frac{\sqrt{\left\langle v^2\right\rangle}}{c} \) \cite{Hogg:1999ad, Muttoni:2023prw}, which arises from the peculiar velocity of the galaxy and accounts for the uncertainty in the redshift measurement. Here we set \( \sqrt{\left\langle v^2\right\rangle} = 500 \, \mathrm{km\,s}^{-1} \). The terms \( \delta(\alpha - \alpha_s) \) and \( \delta(\beta - \beta_s) \) are $\delta$-functions representing the measured sky coordinates, assuming no uncertainties in their measurements.
For dark sirens, since there is no EM signal to provide EM data, this item is treated as a constant and carries no information.}

$p_0(z,\alpha, \beta)$ in eq.~(\ref{con:likelihood}) represents the prior distribution of galaxies in the Universe. {For bright sirens, we have \( p_0(z) \propto z^2 \), assuming that galaxies are uniformly distributed in the comoving volume.
For dark sirens, $p_0(z,\alpha, \beta)$ can be expressed as}
\begin{equation}\label{DEM}
p_0(z,\alpha, \beta)=\frac{1}{N_{\rm in}}\sum_{j=1}^{N_{\rm in}} \mathcal{N}\left(z_j, \sigma_{z, j}\right) \delta(\alpha-\alpha_j)\delta(\beta-\beta_j),
\end{equation}
where $\mathcal{N}\left(z_j, \sigma_{z, j}\right)$ represents a Gaussian distribution centered at $z_j$, with a standard deviation $\sigma_{z, j}$ arising from the peculiar velocity of the $j$-th galaxy and we take
$\sigma_{z} = (1+z) \frac{\sqrt{\left\langle v^2\right\rangle}}{c}$ \cite{Hogg:1999ad, Muttoni:2023prw}. Under the assumption of $\sqrt{\left\langle v^2\right\rangle}=500 \mathrm{~km} \mathrm{~s}^{-1}$, $\sigma_{z}$ is approximately 0.0017 in the redshift range [0, 0.05] \cite{Henriques:2011xn}.
We define $N_{\rm in}$ as the number of potential host galaxies within the localization volume. Generally, a small value of $N_{\rm in}$ indicates a strong ability to localize the GW source. 
Here, we assign an equal weight to each galaxy in the localization volume for simplicity. 
A more rigorous approach is to assign different weights to different galaxies, for instance, replacing $1/N_{\rm in}$ with $\omega_j$, which represents the weight of the $j$-th galaxy and is proportional to the stellar or star-forming luminosity \cite{fishbach2019standard}.

$\gamma (H_0)$ in eq.~(\ref{con:likelihood}) is a normalization factor introduced to account for the selection effects in both GW and EM observations, which can be expressed as
\begin{equation}
\begin{aligned}
\gamma (H_0) = \int \mathcal{D}_{\rm det}^{\rm GW}(d_{\rm L}(z,H_0),\alpha, \beta)\mathcal{D}_{\rm det}^{\rm EM}(z,\alpha, \beta)p_0(z, \alpha, \beta){\rm d}z {\rm d}\alpha {\rm d}\beta,
\end{aligned}
\end{equation}
where $\mathcal{D}_{\rm det}^{\rm GW}(d_{\rm L}(z,H_0),\alpha, \beta)$ and $\mathcal{D}_{\rm det}^{\rm EM}(z,\alpha, \beta)$ represent the GW detection probability and the EM detection probability, respectively. {It should be emphasized that the selection effect term is necessary for both bright and dark sirens. Neglecting this term would introduce a selection bias~\cite{Gair:2022zsa,Gray:2019ksv}.} 

The GW detection probability \cite{Gray:2019ksv} can be expressed as
\begin{align}
\mathcal{D}_{\rm det}^{\rm GW}(d_{\rm L}(z,H_0), \alpha, \beta)=\int_{\mathcal{D}_{\rm GW}>\mathcal{D}_{\rm GW}^{\rm th}}p(\mathcal{D}_{\rm GW}|d_{\rm L}(z,H_0), \alpha, \beta){\rm d}\mathcal{D}_{\rm GW},
\end{align}
where $\rho_{\rm th}=10$ represents the threshold of SNR. 

{As for the EM detection probability, we assume that all the EM counterparts and host galaxies could be detected up to a certain maximum redshift, $z_{\rm max}$. Therefore, the EM detection probability \cite{Chen:2017rfc} is isotropic and depends solely on redshift, which can be expressed as
\begin{equation}\label{pemdet}
    \mathcal{D}_{\rm det}^{\rm EM}(z)
    =\int_{\mathcal{D}_{\rm EM}>\mathcal{D}_{\rm EM}^{\rm th}}p(\mathcal{D}_{\rm EM}|z){\rm d}\mathcal{D}_{\rm EM}\propto\mathcal{H}(z_{\rm max}-z),
\end{equation}
where $\mathcal{H}$ is the Heaviside step function. For bright sirens, we assume that SNRs of EM signals are much higher than those of GW signals within the redshift range detected by the Sloan Digital Sky Survey (SDSS) (see Sec.~\ref{sub:bright}). Therefore, we expect that the impact of EM selection effects is negligible ($z_{\rm max}\to\infty$). 
For dark sirens, the value of $z_{\rm max}$ depends on the maximum redshift covered by our selected galaxy catalog. In this paper, the galaxies we consider are located at $z<0.05$ (see Sec.~\ref{sub:dark}), and therefore we have $z_{\rm max}=0.05$.}

Using eqs.~(\ref{con:likelihood})--(\ref{pemdet}), we can calculate the likelihood of a single GW event. The total likelihood of SMBHB events can be expressed as
\begin{align}
    p_{\rm total}(\mathcal{D}_{\rm GW},\mathcal{D}_{\rm EM}|H_0)=\prod_{k=1}^{N_{\rm SMBHB}} {p(\mathcal{D}_{{\rm GW},k},\mathcal{D}_{{\rm EM},k}|H_0)},
\end{align}
where $N_{\rm SMBHB}$ is the total number of SMBHB events and $k$ represents the $k$-th GW event.

\section{SMBHB bright sirens}
\label{sub:bright}
\begin{figure}
\centering
\includegraphics[angle=0, width=10cm]{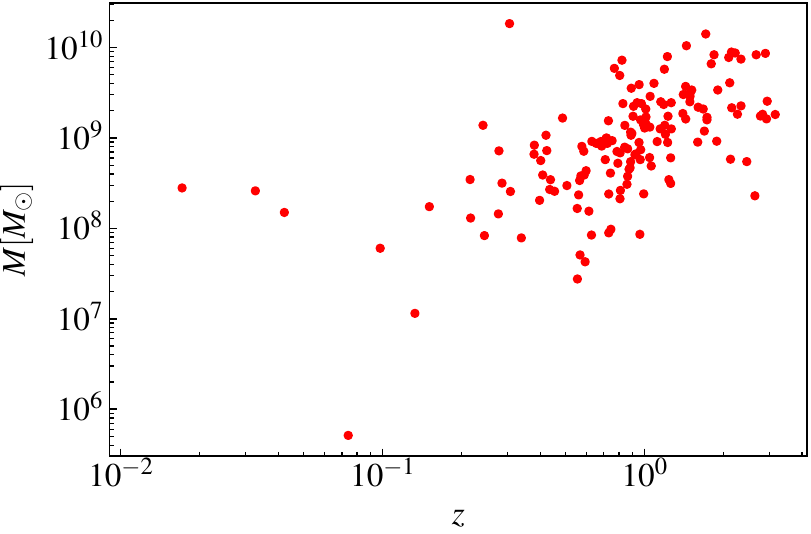}
\caption{\label{fig_bright_zM}
  Distribution of 154 SMBHB candidates in the $z$-$M$ plane, taken from Refs.~\cite{Yan:2015mya,Li:2016hcm,Valtonen:2008tx,Zheng:2015dij,Li:2017eqf,Graham:2015gma,Graham:2015tba,Charisi:2016fqw}. These SMBHB candidates are used in the analysis of bright sirens.
  }
\end{figure}

When simulating the GW bright siren data, we utilize 154 currently available SMBHB candidates obtained from various characteristic signatures in their light curves \cite{Valtonen:2008tx,Graham:2015tba,Charisi:2016fqw} from the Catalina Real-time Transient Survey (CRTS) and the Palomar Transient Factory (PTF) \cite{NANOGrav:2019tvo}. The redshifts of these SMBHB candidates are taken from Refs.~\cite{Yan:2015mya,Li:2016hcm,Valtonen:2008tx,Zheng:2015dij,Li:2017eqf,Graham:2015gma,Graham:2015tba,Charisi:2016fqw}. 
Most of these redshifts are derived from the spectroscopic measurements of SDSS \cite{Paris:2016xdm}. The redshift errors from spectroscopic measurements are negligible compared to the uncertainties introduced by the peculiar velocities of galaxies. Therefore, we do not consider redshift measurement errors in the analyses of bright sirens. Figure~\ref{fig_bright_zM} displays these SMBHB candidates in the $z$-$M$ plane. We use their redshifts to calculate their luminosity distances based on the $\Lambda$CDM model in which the cosmological parameters are set to the \emph{Planck} 2018 results.

According to our calculations, the closest two SMBHB candidates are 4.293 degrees apart in the sky, and the localization precisions of these two candidates by a PTA ($N_{\rm p}=100$ and $\sigma_t=20$ ns) are 0.961 degrees and 0.365 degrees respectively. The localization precisions of the two candidates are evidently smaller than the distance between them. Therefore, a GW source will not have more than one SMBHB candidate in its localization volume. That is to say, if a GW source's localization volume contains a SMBHB candidate, we can be certain that this candidate is the host galaxy of the GW source. SMBHBs detected by PTAs may not necessarily be among the SMBHB candidates. If we can observe the EM signals emitted by SMBHBs, we could still use them as bright sirens. Even if we cannot observe their EM counterparts, we could use them as dark sirens.

\begin{figure}
\centering
\includegraphics[angle=0, width=10.0cm]{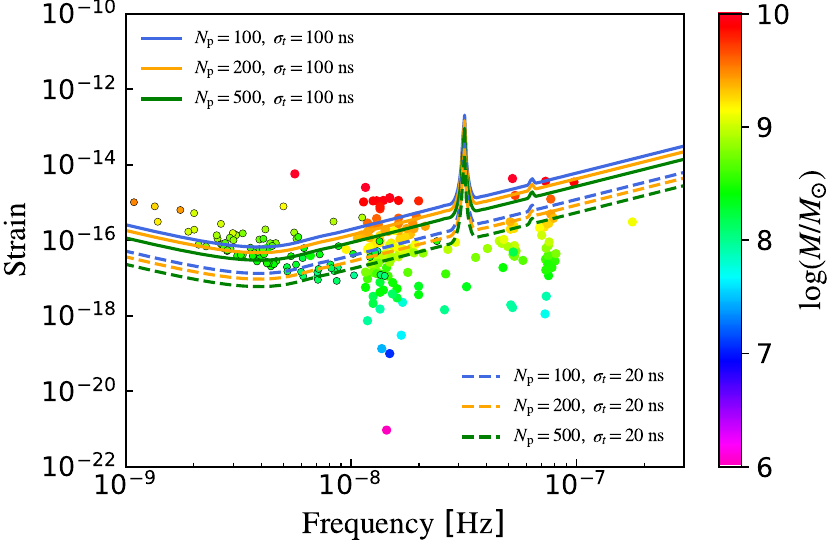}
\caption{\label{fig3}
  Detection curves of SKA-era PTAs with a 10-year observation time span. The solid and dotted lines represent the cases of $\sigma_t=100$ ns and $\sigma_t=20$ ns, respectively. The data points represent the GW strain amplitudes ($h_0$) when $f=f_0$, with $h_0=2[M_{\rm c}(1+z)]^{5/3} (\pi f)^{2/3} d_{\rm L}^{-1}$ \cite{Yan:2019sbx} and $f_0$ the GW frequency at the time of the first observation. The solid dots without black borders represent 154 SMBHB candidates, and the solid dots with black borders represent 84 SMBHBs simulated from the 2MASS catalog.
  }
\end{figure}

We use eqs.~(\ref{s-definition})--(\ref{ft}) to simulate the GW signals emitted by these SMBHB candidates, and plot the detection curves of SKA-era PTAs (averaged over the sky locations of GW sources) in figure.~\ref{fig3} by using the {\tt hasasia} package \cite{Hazboun:2019nqt, Hazboun:2019vhv}. Here we only plot the strain amplitudes of GW signals when $f=f_0$, because a simple calculation using eq.~(\ref{ft}) shows that the variation of the GW frequency of an inspiraling SMBHB with $M=10^9$ $M_\odot$ and $f_0=10^{-7}$ Hz in a 10-year observational time span is $4.36\times 10^{-9}$ Hz. This variation is so minuscule that the amplitude of the GW strain undergoes only negligible changes over the given time span. The solid dots without black borders represent 154 SMBHB candidates \cite{Graham:2015gma,Graham:2015tba,Charisi:2016fqw,Yan:2015mya,Li:2016hcm,Valtonen:2008tx,Zheng:2015dij,Li:2017eqf}.
As $N_{\rm p}$ increases and $\sigma_t$ decreases, the more sensitive detection curves enable ones to detect more SMBHBs. The dotted curves ($\sigma_t=20$ ns) are obviously lower than the solid curves ($\sigma_t=100$ ns), indicating that the rms of timing residual has a more dominating effect than the number of MSPs on the detections of SMBHBs.


\begin{figure}
\centering
\includegraphics[angle=0, width=10.0cm]{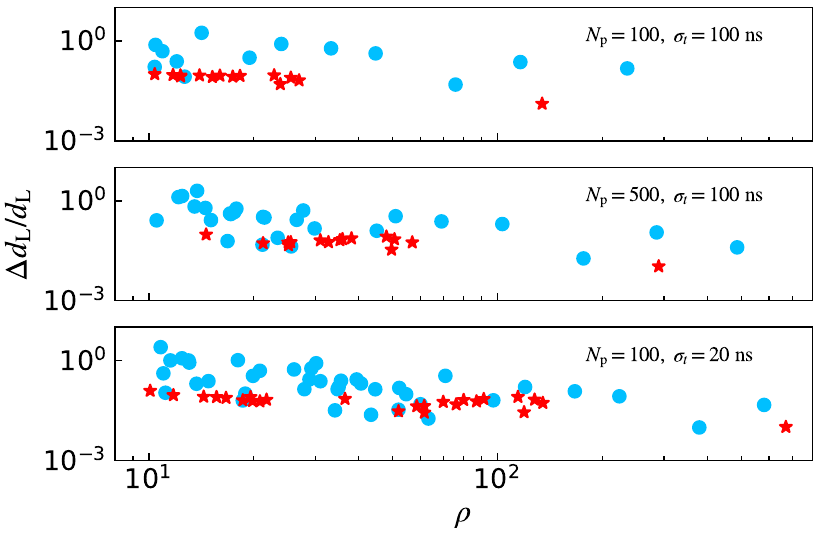}
\caption{\label{fig4}
  Measurement precision of luminosity distance ($\Delta d_{\rm L}/d_{\rm L}$) as a function of SNR ($\rho$). The red stars and the blue dots represent the detected SMBHBs with $\rho >$ 10, used as the bright and dark sirens, respectively. 
  The impacts of $N_{\rm p}$ and $\sigma_t$ on the detections of SMBHB can be explicitly seen.
 }
\end{figure}

The $1\sigma$ relative errors of the luminosity distances ($\Delta d_{\rm L}/d_{\rm L}$) of the mock SMBHB bright sirens as a function of SNR, $\rho$, are shown in figure~\ref{fig4}. The corresponding numbers of detected bright ($\rho>10$) are represented by $N_{\rm s}$ and shown in table~\ref{table1}.
In the case of $N_{\rm p}=100$, the number of detected bright sirens increases from 14 ($\sigma_t=100$ ns) to 25 ($\sigma_t=20$ ns).
Although the number of MSPs can also affect the detection of SMBHBs, its effect is not as obvious as $\sigma_t$.
For example, in the case of $\sigma_t=100$ ns, the number of detected bright sirens increases only from 14 ($N_{\rm p}=100$) to 15 ($N_{\rm p}=500$).
This indicates that the rms of timing residual is the most significant factor affecting the number of bright sirens and the errors of luminosity distances. Our results show that approximately 100 MSPs are adequate for detecting individual SMBHBs, provided that the timing measurements can achieve a high enough precision.

\begin{figure}
\centering
\includegraphics[angle=0, width=10cm]{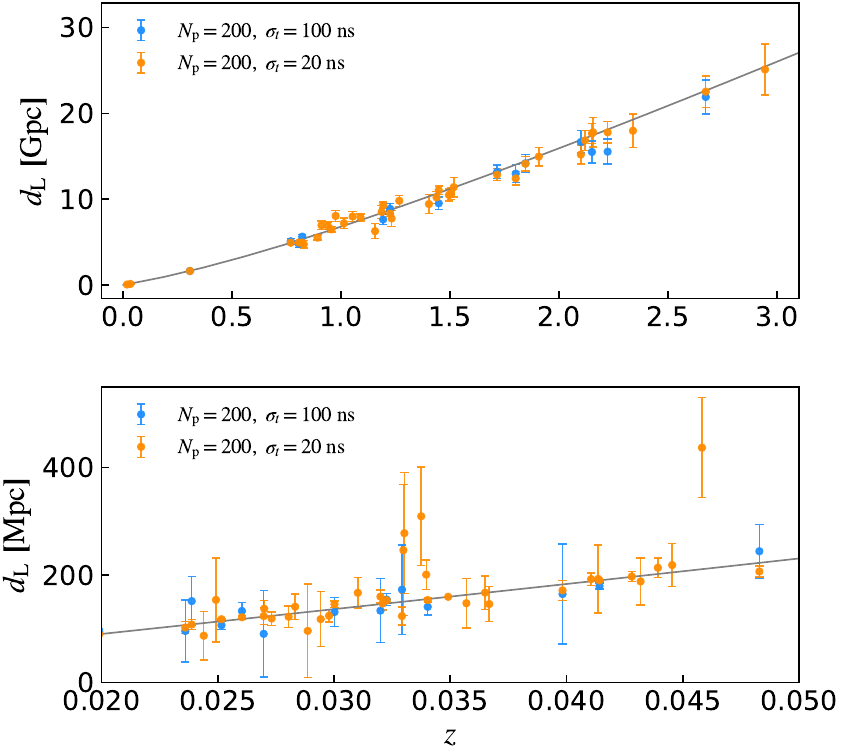}
\caption{\label{fig5}
  GW bright siren and dark siren data simulated from the 154 SMBHB candidates and 84 mock SMBHBs, respectively.
  Upper and lower panels correspond to the bright and dark sirens, respectively.
  The redshifts of dark sirens shown in the lower panel represent the redshifts of mock SMBHBs' host galaxies.
  The luminosity distances $d_{\rm L}$ are calculated based on the $\Lambda$CDM model, with the fiducial values of cosmological parameters set to the \emph{Planck} 2018 results. The 1$\sigma$ error bars of the data points ($\Delta d_{\mathrm{L}}$) in the figure are obtained from the Fisher matrix. The central values of the data points are randomly chosen in the range of [$d_{\rm L}-\Delta d_{\mathrm{L}}$, $d_{\rm L}+\Delta d_{\mathrm{L}}$]. The data points with $\Delta d_{\mathrm{L}}/d_{\rm L}>1$ are not displayed in the figure.
  }
\end{figure}

Here we consider the base $\Lambda$CDM model ($w=-1$) and the $w$CDM model ($w={\rm constant}$). The $d_{\rm L}$-$z$ relation can be expressed as
\begin{align}
    d_{\rm L} = \frac{(1+z)}{H_0} \int^{z}_{0} \frac{ {\rm d} z^{\prime}}{\sqrt{\Omega_{\rm m}(1+z^{\prime})^3 + (1-\Omega_{\rm m})(1+z^{\prime})^{3(1+w)}}},
\end{align}
where $\Omega_{\rm m}$ represents the current matter density parameter.
Assuming different values of  $N_{\rm p}$ and $\sigma_t$, we simulate six sets of bright siren data that include $d_{\rm L}$, $\Delta d_{\rm L}$, and redshift $z$ of the SMBHB candidates. Two representative sets of the GW bright siren data are shown in the upper panel of figure~\ref{fig5}. The numbers of detected bright sirens in the case of $\sigma_{t}=20$ ns are much larger than those in the case of $\sigma_{t}=100$ ns for the same number of MSPs. Decreasing $\sigma_t$ improves SNRs of GW events and reduces the measurement errors of luminosity distances. We use these bright siren data to constrain the $\Lambda$CDM and $w$CDM models, respectively.

\begin{figure}
\centering
\includegraphics[angle=0, width=10cm]{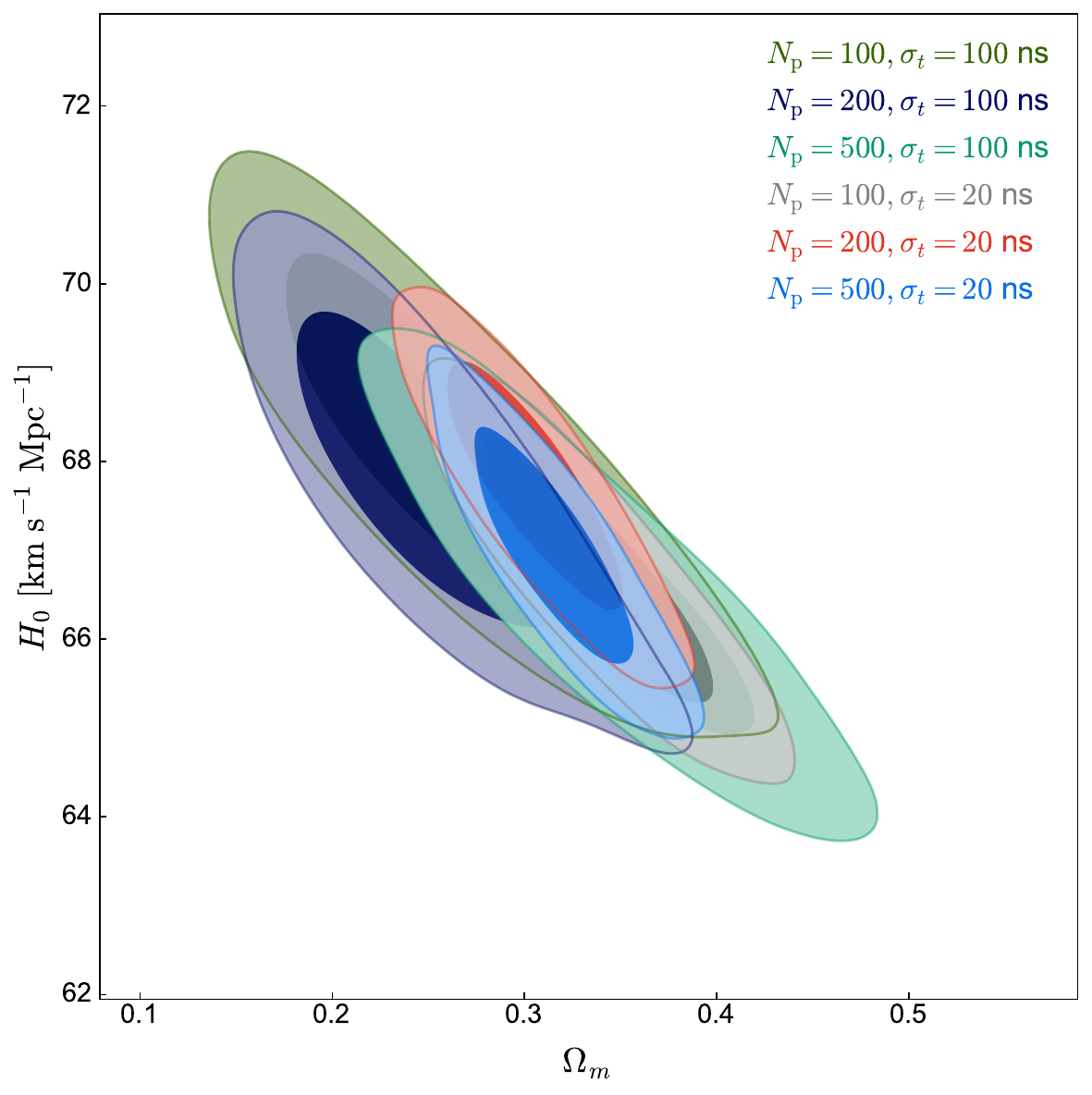}
\caption{\label{fig_LCDM}
  2D marginalized contours (68.3\% and 95.4\% confidence level) in the $\Omega_{\rm m}$-$H_0$ plane for the $\Lambda$CDM model using mock GW bright siren data.
  }
\end{figure}

The constraint results of the base $\Lambda$CDM model are shown in figure~\ref{fig_LCDM} and listed in table~\ref{table1}. 
We define the constraint precision of the parameter $\xi$ as $\varepsilon(\xi) = \sigma(\xi)/\xi$ with $\sigma(\xi)$ representing the marginalized absolute error. In the case of $\sigma_t=100$ ns, as $N_{\rm p}$ increases from 100 to 500, $\varepsilon(H_0)$ decreases from 1.4\% to 1.1\%.
In the case of $N_{\rm p}=100$, as $\sigma_t$ decreases from 100 ns to 20 ns, $\varepsilon(H_0)$ decreases from 1.4\% to 0.98\%. We note that reducing $\sigma_t$ is more effective than increasing $N_{\rm p}$ on improving the constraining capability of bright sirens. If $\sigma_t$ could reach 20 ns, 100 MSPs would be adequate to achieve a measurement precision for $H_0$ comparable to that of the current cosmic distance-ladder observation.

\begin{figure}
\centering
\includegraphics[angle=0, width=10.0cm]{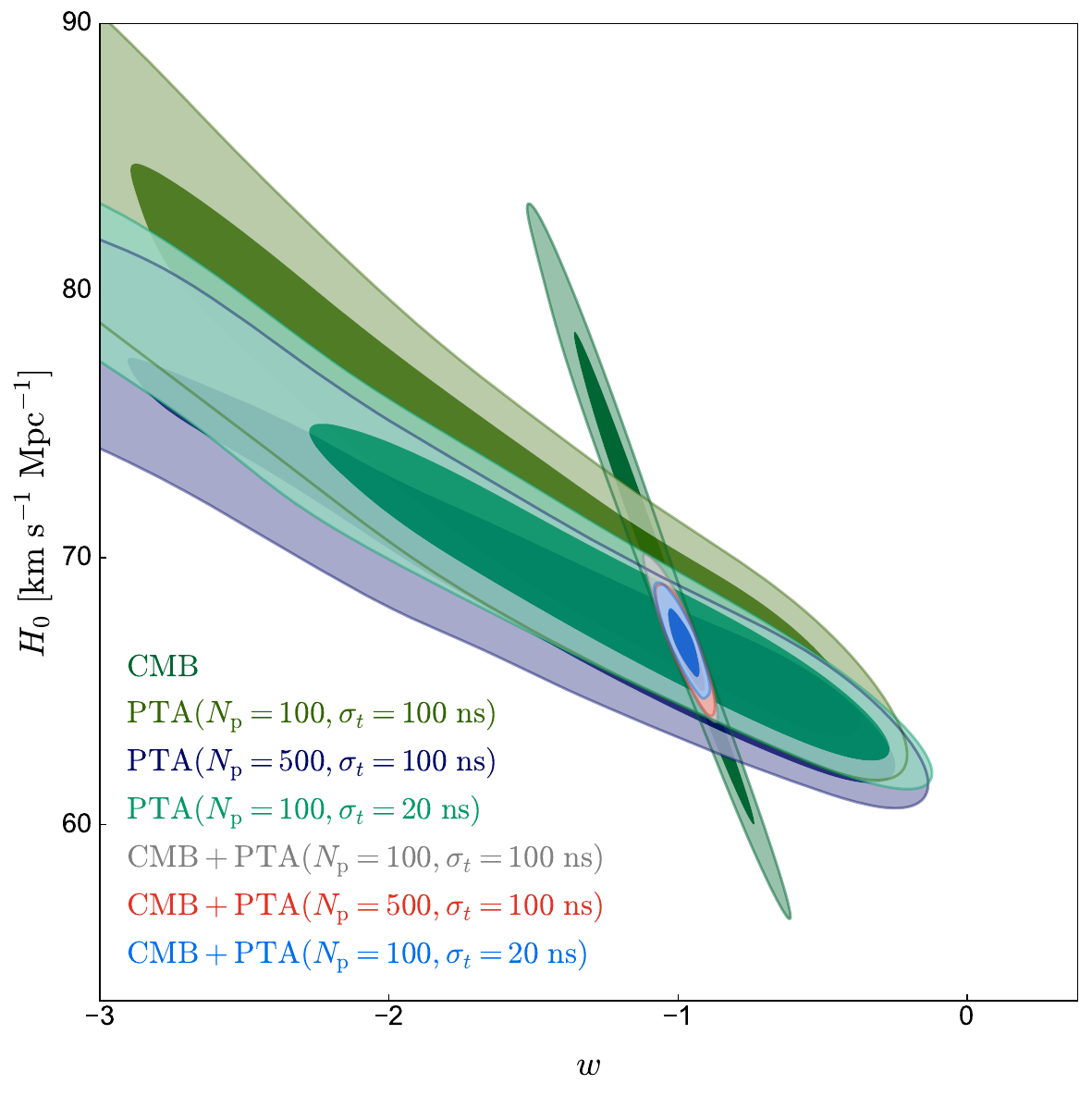}
\caption{\label{fig7}
  2D marginalized contours (68.3\% and 95.4\% confidence level) in the $w$-$H_0$ plane for the $w$CDM model using the CMB, PTA, and CMB+PTA data. Here, the PTA data refer to the mock GW bright siren data. 
}
\end{figure}

The constraint results of the $w$CDM model are shown in figure~\ref{fig7} and listed in table~\ref{table2}.
In the $w$CDM model, the cosmic microwave background (CMB) data cannot provide tight constraints on the EoS parameter of dark energy ($w$), because CMB encodes information from the early Universe, whereas dark energy becomes dominant in the late Universe. Nevertheless, figure~\ref{fig7} shows that the CMB data and the bright siren data (simply referred to as the PTA data) have distinct degeneracy orientations in the $w$-$H_0$ plane. This indicates that while the PTA data alone cannot constrain $w$ effectively, it can provide tight constraints on $H_0$, thereby breaking the degeneracy between the parameters $w$ and $H_0$. Table~\ref{table2} shows that, in the case of $N_{\rm p}=100$ and $\sigma_{t}=20$ ns, the combination of the CMB and PTA data gives the relative error $\varepsilon(w)=3.7\%$, which is roughly comparable with the result of \emph{Planck} 2018 TT,TE,EE+lowE+lensing+SNe+BAO \cite{Planck:2018vyg}. The results suggest that the SMBHB bright sirens will be a useful probe to explore the nature of dark energy.

\begin{table}
\centering
\setlength\tabcolsep{12pt}
\renewcommand\arraystretch{1.5}
\begin{tabular}{cccccccccc}
\hline\hline
& & & \multicolumn{3}{c}{\bf Bright siren} & & \multicolumn{3}{c}{\bf Dark siren} \\
\cline{4-6} \cline{8-10}
$N_{\rm p}$ & $\sigma_{t} ({\rm ns})$ &
& $N_{\rm s}$ & $\sigma(H_{\rm 0})$ & $\varepsilon(H_{\rm 0})$ &
& $N_{\rm s}$ & $\sigma(H_{\rm 0})$ & $\varepsilon(H_{\rm 0})$ \\ \hline
100 & 100 &   & 14 & 1.4  & 0.0206  & & 13 & 3.458 & 0.0500   \\
200 & 100 &   & 14 & 1.2  & 0.0177  & & 19 & 2.350 & 0.0339   \\
500 & 100 &   & 15 & 1.1  & 0.0166  & & 27 & 1.824 & 0.0270   \\
100 & 20  &   & 25 & 0.98  & 0.0147  & & 41 & 1.248 & 0.0184   \\
200 & 20  &   & 40 & 0.90  & 0.0133  & & 49 & 1.066 & 0.0159   \\
500 & 20  &   & 53 & 0.88  & 0.0131  & & 56 & 0.895 & 0.0131   \\
\hline\hline
\end{tabular}
\caption{\label{table1}
Relative errors of $H_0$ in the $\Lambda$CDM model.
The GW bright siren data are simulated based on the 154 SMBHB candidates, while the GW dark siren data are simulated based on the 84 mock SMBHBs from the 5119 galaxies in the 2MASS catalog.
$N_{\rm s}$ represents the number of detected SMBHBs ($\rho >10$) and $\varepsilon(H_{\rm 0})$ represents the relative error of $H_0$. 
}
\end{table}

\begin{table}
\setlength\tabcolsep{10pt}
\renewcommand\arraystretch{1.5}
\centering
\begin{tabular}{llccccccc}
\hline\hline
Data  &$N_{\rm p}$ & $\sigma_{t} ({\rm ns})$
& $\sigma(\Omega_{\rm m})$ & $\varepsilon(\Omega_{\rm m})$
& $\sigma(H_{\rm 0})$ & $\varepsilon(H_{\rm 0})$
& $\sigma(w)$ & $\varepsilon(w)$ \\
\hline
    CMB                  & $-$ & $-$ &   0.054  & 0.179  & 6.0 & 0.858  & 0.20 & 0.185   \\ \hline
\multirow{3}{*}{PTA}     & 100 & 100 &   0.067  & 0.260  & 6.1 & 0.084  & 0.86 & 0.521    \\ \cline{2-9}
                         & 500 & 100 &   0.052  & 0.173  & 4.4 & 0.063  & 0.76 & 0.539   \\ \cline{2-9}
                         & 100 & 20  &   0.044  & 0.125  & 4.1 & 0.059  & 0.68 & 0.511   \\ \hline
\multirow{3}{*}{CMB+PTA} & 100 & 100 &   0.012  & 0.037  & 1.3 & 0.019  & 0.049 & 0.050   \\ \cline{2-9}
                         & 500 & 100 &   0.010  & 0.031  & 1.0 & 0.015  & 0.042 & 0.043   \\ \cline{2-9}
                         & 100 & 20  &   0.009  & 0.026  & 0.9 & 0.013  & 0.037 & 0.037   \\
\hline\hline
\end{tabular}
\caption{\label{table2}
  Relative errors of the cosmological parameters in the $w$CDM model using the CMB, PTA, and CMB+PTA data.
  $N_{\rm s}$ is the number of detected SMBHB ($\rho>10$). Here, the PTA data refer to the mock GW bright siren data.
}
\end{table}

\begin{figure}
\centering
\includegraphics[angle=0, width=10cm]{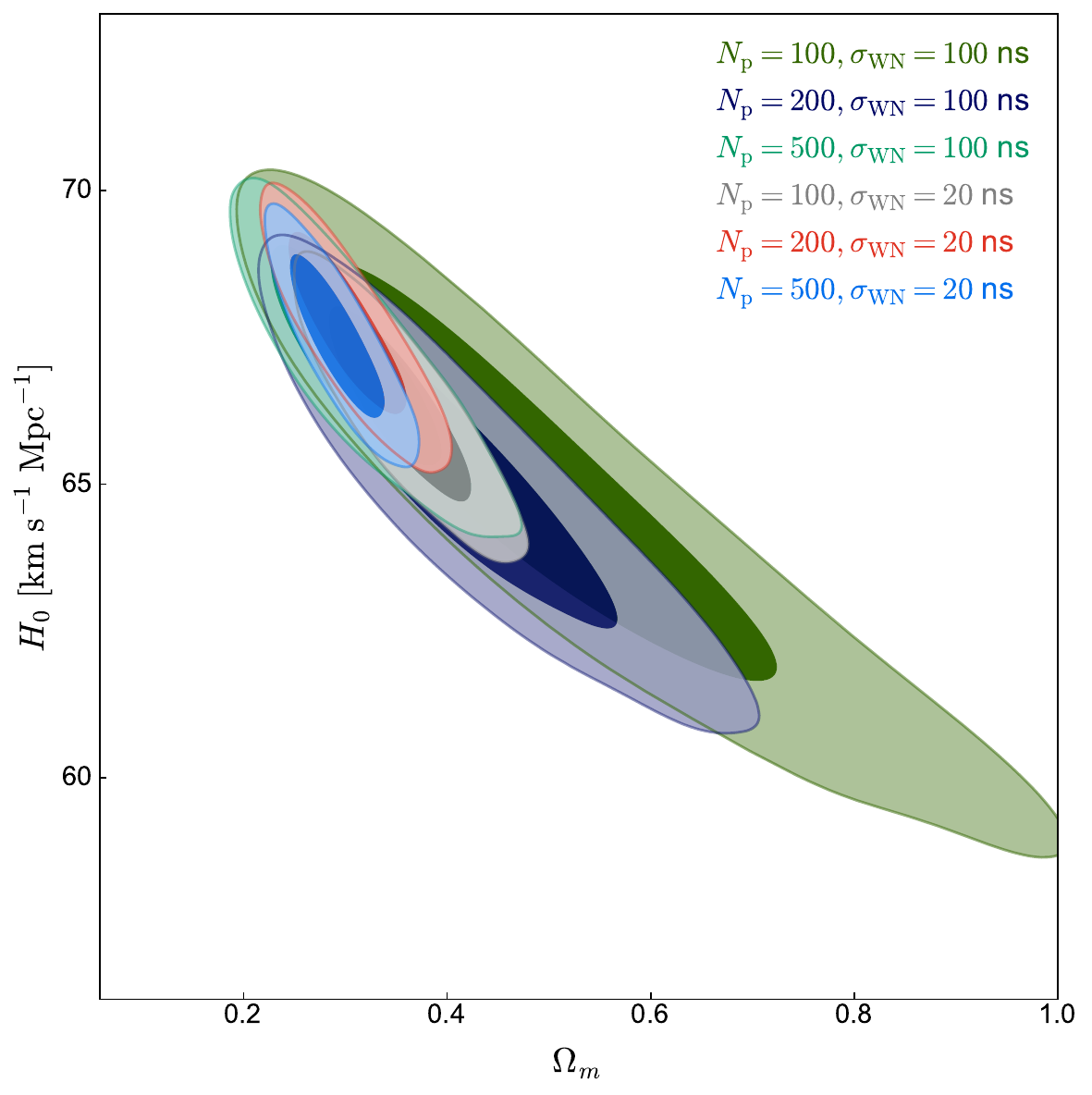}
\caption{\label{fig8}
Same as figure~\ref{fig_LCDM} except considering the
impact of the intrinsic red noise and the DM variations noise. Here, $\sigma_{\rm WN}$ represents rms of white noise.
}
\end{figure}

To demonstrate the effect of red noise, we analyze additional scenarios where we take into account the intrinsic red noise (IRN) \cite{Shannon:2010bv} and the dispersion measurements (DM) variations noise \cite{Keith:2012ht}. The power spectral densities of the intrinsic red noise and the dispersion measurements variations noise are expressed as \cite{Antoniadis:2022pcn}
\begin{align}
\label{rn_1}
&P_{\mathrm{IRN}}(f)=\frac{A^2_{\mathrm{IRN}}{ }}{12 \pi^{2}} f_{\mathrm{yr}}^{-3}\left(\frac{f}{f_{\mathrm{yr}}}\right)^{-\gamma_{\mathrm{IRN}}}, \\
&P_{\mathrm{DM}}(f, \nu)=\frac{A^2_{\mathrm{DM}}{ }}{12 \pi^{2}} f_{\mathrm{yr}}^{-3}\left(\frac{f}{f_{\mathrm{yr}}}\right)^{-\gamma_{\mathrm{DM}}}\left(\frac{1400 ~ \mathrm{MHz}}{{\nu}}\right)^{2},
\end{align}
respectively, with $f_{\rm yr} = \rm{yr}^{-1}$. Here, $\nu$ represents the radio observation frequency and is set to $1400~{\rm MHz}$, thereby $P_{\mathrm{DM}}(f, \nu)$ expressed as $P_{\mathrm{DM}}(f)$. $A_{\rm IRN/DM}$ and $\gamma_{\rm IRN/DM}$ are the amplitude parameter and the spectral index, respectively, unique to each MSP. Here, we obtain the values of $A_{\rm IRN/DM}$ and $\gamma_{\rm IRN/DM}$ from the IPTA second data release \cite{Antoniadis:2022pcn}.
Following the methods in Ref.~\cite{Siemens:2013zla}, the timing residuals induced by red noise can be expressed as
\begin{equation}
\sigma_{\rm IRN/DM}^{2}=2 \int_{f_{\rm L}}^{f_{\rm H}} P_{\mathrm{IRN/DM}}(f) \mathrm{~d} f~ = \frac{(1-(N_t/2)^{-\gamma_{\rm IRN/DM}+1})(2b {T_{\rm obs}}^{\gamma_{\rm IRN/DM}-1})}{\gamma_{\rm IRN/DM}-1},
\label{sigma_red}
\end{equation}
with $f_{\rm L}=1/{T_{\rm obs}}$, $f_{\rm H}=N_t/2{T_{\rm obs}}$, and $b=P_{\rm{IRN/DM}}(f)/f^{-\gamma_{\rm IRN/DM}}$. The cadence of monitoring the pulses from MSPs is set to two weeks, while the observation span $T_{\rm obs}$ is 10 years. The total rms of red noise can be expressed as
\begin{align}
\sigma_{\rm RN} = \sqrt{{\sigma_{\rm IRN}^2}+{\sigma_{\rm DM}^2}}.
\label{total_red}
\end{align}
We use eqs.~(\ref{rn_1})--(\ref{total_red}) to calculate $\sigma_{\rm RN}$ and select 10 MSPs with the smallest $\sigma_{\rm RN}$ from the IPTA second data release. The 10 smallest $\sigma_{\rm RN}$ values are 0.01 ns, 0.14 ns, 0.22 ns, 8.70 ns, 20.12 ns, 91.63 ns, 307.05 ns, 314.80 ns, 406.35 ns, and 432.38 ns, respectively. In the SKA era, we assume that a sufficient number of MSPs with a similar red-noise level to these 10 MSPs could be observed. For each MSP we consider, we randomly select a $\sigma_{\rm RN}$ value from the 10 smallest values and add it to the rms of the considered MSP.
The total rms of each MSP is expressed as
\begin{align}
\sigma_t = \sqrt{{\sigma_{\rm WN}^2} + {\sigma_{\rm RN}^2}},
\label{total_rms}
\end{align}
where $\sigma_{\rm WN}$ represents the total rms of white noise. Based on the methods described above, we simulate the bright siren data by incorporating red noise. Subsequently, we use these data to constrain the $\Lambda$CDM model. The constraint results are shown in figure~\ref{fig8} and table~\ref{table4}. 
The addition of red noise worsens the constraint precision of $H_0$ by 1.1 -- 1.7 times in the case of $\sigma_t=100$ ns, but by 1.0 -- 1.1 times in the case of $\sigma_t=20$ ns.
This is because lower white noise leads to higher precision of $H_0$, making the impact of red noise less noticeable. Additionally, compared with red noise, white noise can be more easily reduced through advancements in detection technology. This allows us to enhance the capability of SMBHB standard sirens by minimizing white noise as much as possible. It should be noted that if the effect of SGWB is considered, it is equivalent to considering larger red noise, which would make the constraint precision of $H_0$ further worse. We will discuss this issue in detail in future works.

\begin{table}
\centering
\setlength\tabcolsep{12pt}
\renewcommand\arraystretch{1.5}
\begin{tabular}{cccccccccc}
\hline\hline
$N_{\rm p}$ & $\sigma_{\rm{WN}} ({\rm ns})$ &
& $N_{\rm s}$ & $\sigma(H_{\rm 0})$ & $\varepsilon(H_{\rm 0})$ \\ \hline
100 & 100 &   & 10 & 2.4  & 0.0362   \\
200 & 100 &   & 11 & 1.8  & 0.0269   \\
500 & 100 &   & 15 & 1.2  & 0.0179   \\
100 & 20  &   & 20 & 1.1  & 0.0166   \\
200 & 20  &   & 38 & 1.0  & 0.0148   \\
500 & 20  &   & 50 & 0.91  & 0.0135   \\
\hline\hline
\end{tabular}
\caption{\label{table4}
Relative errors of $H_0$ in the $\Lambda$CDM model using mock GW bright siren data and considering the intrinsic red noise and the DM variations noise. 
These data are simulated based on the 154 SMBHBs candidates. Here, $\sigma_{\rm WN}$ represents rms of white noise.
}
\end{table}

\section{SMBHB dark sirens}
\label{sub:dark}
{To simulate SMBHB dark sirens, we need to select a suitable galaxy catalog. The mass of an SMBHB is related to the mass of its host galaxy, which can be estimated based on the galaxy's absolute magnitude. Estimating the absolute magnitude requires information on the galaxy's apparent magnitude and luminosity distance. In Ref.~\cite{Mingarelli:2017fbe}, the authors specifically estimated the luminosity distances of galaxies in the 2 Micron All Sky Survey (2MASS) catalog by correcting for the peculiar velocities of galaxies. Therefore, we select the 2MASS catalog for simulating the dark siren data. The galaxies with $K$-band absolute magnitudes less than -25 can be considered as the candidate host galaxies for SMBHBs \cite{Rosado:2013wva}. Based on this criterion, we select 5119 galaxies from the 2MASS catalog.}
These galaxies are distributed in the local Universe ($z < 0.05$), and the 2MASS catalog can be considered complete in this redshift range \cite{Mingarelli:2017fbe}. The mass distribution of these galaxies is in $10^{11}$ -- $10^{12}$ $M_{\odot}$ \cite{Mingarelli:2017fbe,Feng:2020nyw}.
We estimate the masses of SMBHBs in these galaxies based on the $M$-$M_{\rm bulge}$ relationship, where $M_{\rm bulge}$ represents the bulge mass of a galaxy \cite{McConnell:2012hz}.

The probability that a galaxy hosts an SMBHB in the PTA band, $p_j$, can be expressed as
\begin{equation}
p_{j}=\frac{t_{{\rm c}, j}}{T_{\rm life}} \int_{0.25}^{1} \mathrm{~d} \mu_{\star} \frac{\mathrm{d} N}{\mathrm{~d} t}\left(M_{\star}, \mu_{\star}, z'\right) T_{\rm life}.
\label{pj}
\end{equation}
Here, $t_{{\rm c}, j} = (5/256) (\pi f_{\rm low})^{-8/3} (G\mathcal{M}_{\rm c}/c^3)^{-5/3}$ is the time to SMBHB coalescence in the $j$-th galaxy, with $f_{\rm low}= 1$ nHz being the lower limit of the PTA band.
$T_{\rm life}$ is the effective lifetime of an SMBHB \cite{Mingarelli:2017fbe,BinneyTremaine,Sesana:2015haa}.
$\mathrm{d} N / \mathrm{d} t\left(M_{\star}, \mu_{\star}, z'\right)$ is the galaxy merger rate from the Illustris cosmological simulation project \cite{Rodriguez-Gomez:2015aua, Genel:2014lma}, with $M_{\star}$ the stellar masses of the galaxies, $\mu_{\star}$ the progenitor stellar mass ratio, and $z'$ the redshift at which the galaxies merge.
We determine the number of SMBHBs in the galaxy catalog as $N_{\rm SMBHB} = \sum_{j} p_j$ \cite{Rodriguez-Gomez:2015aua, Genel:2014lma}. Eq.~(\ref{pj}) shows that $p_j$ depends on $t_{{\rm c}, j}$, and $t_{{\rm c}, j}$ is related to $\mathcal{M}_{\rm c}$ that is determined by $q$. Therefore, $q$ could affect the merger probability of SMBHBs and consequently influence the number of simulated SMBHBs. As shown in table~\ref{table3}, smaller values of $q_{\rm min}$ increase the numbers of both simulated and detected SMBHBs. As $q_{\rm min}$ decreases from 1 to 0.01, the number of mock SMBHBs increases from 54 to 197. The main results of this paper are based on the assumption of $q_{\rm min}=0.25$. Under this assumption, we find that there are approximately 84 SMBHBs in the total 5119 galaxies. Then we randomly select 84 galaxies from the total number of galaxies as host galaxies for SMBHBs based on the probability distribution $p_j$.

\begin{figure}
\centering
\includegraphics[angle=0, width=10cm]{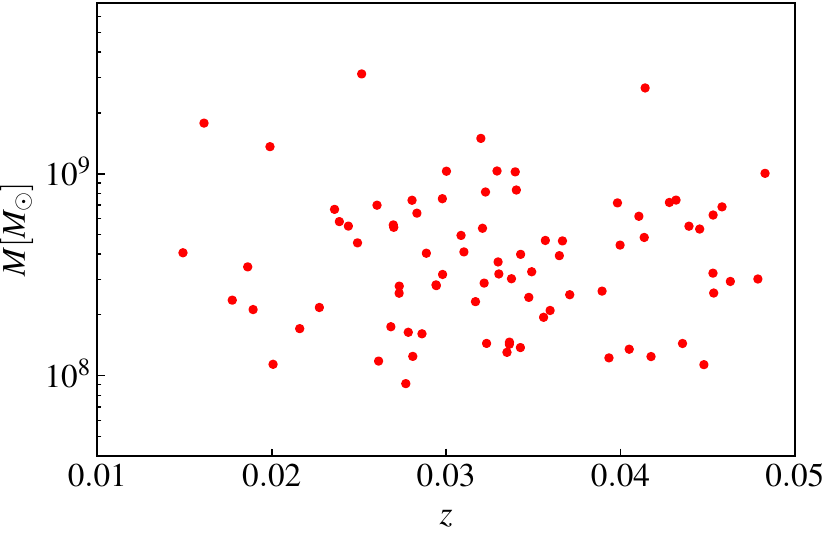}
\caption{\label{fig_dark_zM}
  Distribution of 84 SMBHBs simulated from the 2MASS catalog in the $z$-$M$ plane. We use these SMBHBs in the analysis of dark sirens.
  }
\end{figure}

Figure~\ref{fig_dark_zM} displays the 84 SMBHBs in the $z$-$M$ plane. Figure~\ref{fig3} shows the strain amplitudes (when $f=f_0$) of the GW signals emitted by the 84 SMBHBs, with $f_0$ calculated using the following formula,
\begin{align}\label{f0}
    f_0 = \pi^{-1} \left(\frac{G\mathcal{M}_{\rm c}}{c^3}\right)^{-5/8} \left(\frac{256}{5}t_{\rm c}\right)^{-3/8},
\end{align}
where $t_{\rm c}$ is taken from a uniform distribution in [100 yr, 26 Myr]. The solid dots with black borders represent 84 SMBHBs simulated from the 2MASS \cite{2MASS:2006qir} Extended Source Catalog \cite{Jarrett:2000me}.
The $1\sigma$ relative errors of the luminosity distances ($\Delta d_{\rm L}/d_{\rm L}$) of the mock SMBHB dark sirens as a function of SNR, $\rho$, are shown in figure~\ref{fig4}. The corresponding numbers of detected dark sirens ($\rho>10$) are shown in table~\ref{table1}. In the case of $N_{\rm p}=100$, the number of detected dark sirens increases from 13 ($\sigma_t=100$ ns) to 41 ($\sigma_t=20$ ns).
In the case of $\sigma_t=100$ ns, the number of detected dark sirens increases from 13 ($N_{\rm p}=100$) to 27 ($N_{\rm p}=500$). This indicates that the rms of timing residual is the most important factor affecting the number of dark sirens and the errors of luminosity distances, which is consistent with the analysis of bright sirens in the previous text.

\begin{figure}
\centering
\includegraphics[angle=0, width=10cm]{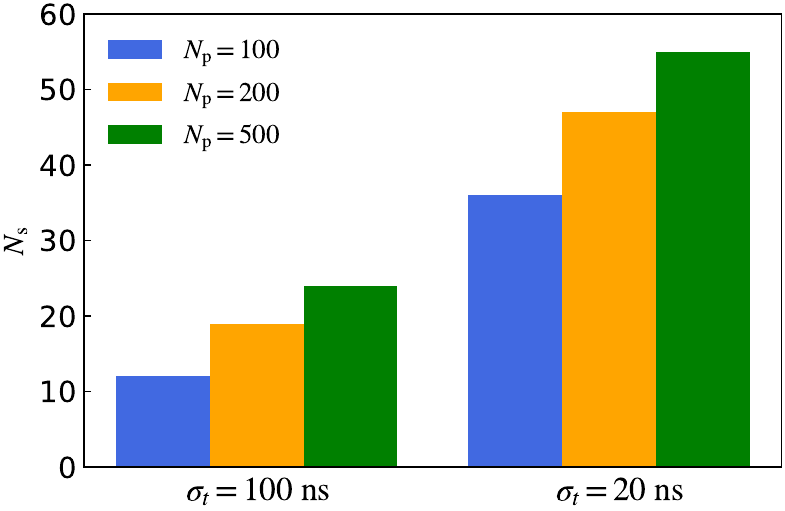}
\caption{\label{fig10}
  Numbers of SMBHBs with $N_{\rm in} <$ 10 in the analysis of dark sirens. $N_{\rm s}$ is the number of detected SMBHBs ($\rho>10$). Here $N_{\rm in}$ is calculated by fixing $H_0=67.36$ $\ksm$. 
  As $N_{\rm p}$ increases and $\sigma_t$ decreases, more SMBHBs satisfy $N_{\rm in}<10$.
}
\end{figure}

Figure~\ref{fig10} shows the numbers of SMBHBs satisfying $N_{\rm in}<10$. It can be seen that more SMBHBs satisfy $N_{\rm in}<10$ as $N_{\rm p}$ increases and $\sigma_t$ decreases. The values of $N_{\rm in}$ depend on the prior values of $H_0$; we fix $H_0=67.36$ $\ksm$ only when plotting figure~\ref{fig10}. In the other parts of this paper, $N_{\rm in}$ still varies with $H_0$. Since the dark sirens are simulated in the local Universe in which the $d_{\rm L}$-$z$ relation shows weak dependence on cosmological models, these data cannot constrain $w$ effectively. Therefore, we only calculate the posterior distribution of $H_0$ using the Bayesian analysis method.

\begin{figure}
\centering
\includegraphics[angle=0, width=10.0cm]{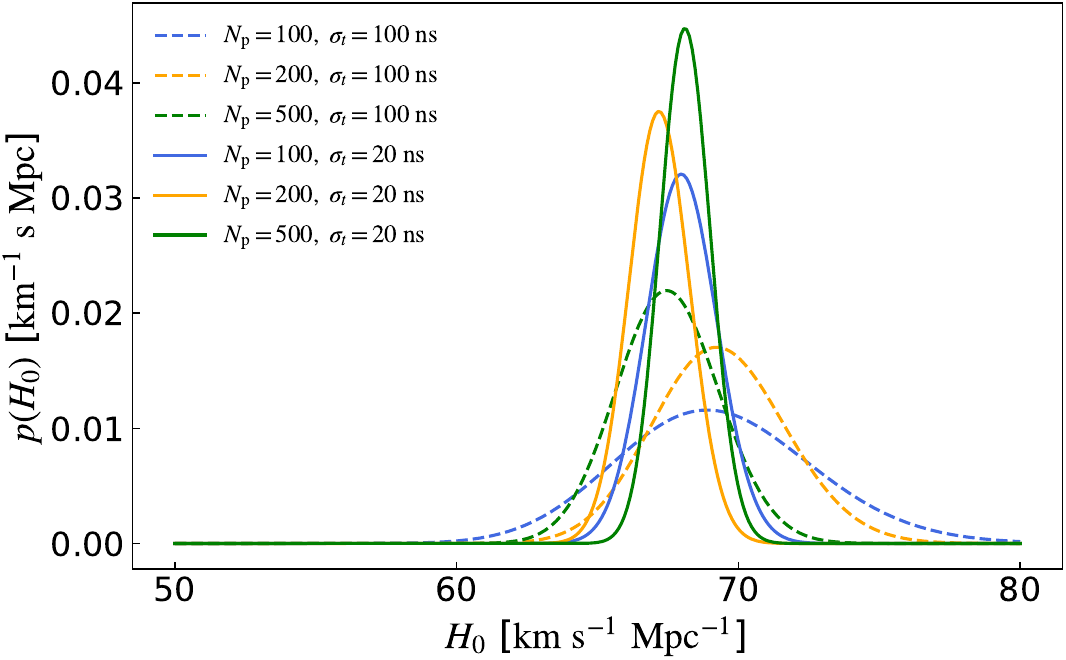}
\caption{\label{fig11}
  1D posterior distribution of $H_0$ inferred from the mock GW dark siren data. The dotted and solid lines represent the cases of $\sigma_t=100$ ns and $\sigma_t=20$ ns, respectively. 
  The errors of $H_0$ decrease as $N_{\rm p}$ increases and $\sigma_t$ decreases.
}
\end{figure}

The results of the SMBHB dark sirens are shown in figure~\ref{fig11} and table~\ref{table1}. In the case of $\sigma_t=100$ ns, increasing $N_{\rm p}$ from 100 to 500 can significantly improve the measurement of $H_0$. The 1$\sigma$ errors of $H_0$ with $\sigma_t=20$ ns are obviously smaller than those with $\sigma_t=100$ ns. It is worth noting that even with only 100 MSPs ($N_{\rm p}$ = 100 and $\sigma_t$ = 20 ns), the precision of $H_0$ could reach $\sim 1.8\%$. Compared with the SMBHB bright sirens, the SMBHB dark sirens have a similar capability of measuring $H_0$. This indicates that even if it is difficult to detect EM counterparts of SMBHBs in the future, dark sirens could solely provide precise measurements of $H_0$. The bright and dark sirens have the potential to complement each other, providing precise measurements for both $w$ and $H_0$.


\section{Discussion}
\label{sec:discussion}

In this work, we assume that the cadence of monitoring the pulses from MSPs is two weeks \cite{Yan:2019sbx} and consider $N_{\rm p}=100,200,500$ respectively.
Actually, observing 500 MSPs is not achievable with this cadence due to the time required for each observation. Therefore, the case of $N_{\rm p}=500$ is used as an extreme scenario for comparison. To demonstrate the impact of the cadence, we consider another scenario with a monthly cadence instead of bi-weekly. In this case, 23 bright sirens could be observed over 10 years when $N_{\rm p}=100$ and $\sigma_{t}=20$ ns. The measurement precision of $H_0$ reaches $1.65\%$, similar to the result obtained with a cadence of two weeks [$\varepsilon(H_0)=1.47\%$]. This indicates that even if the observation time is reduced by a factor of 2, SMBHB standard siren data could still maintain tight constraints on $H_0$.

When simulating GW bright siren data, we utilize 154 currently available SMBHB candidates, primarily identified through the observations of periodic variations in their light curves \cite{Valtonen:2008tx,Graham:2015tba,Charisi:2016fqw} from CRTS and PTF \cite{NANOGrav:2019tvo}. These methods are suitable for SMBHBs in the inspiral phase. Actually, SMBHBs in the merger phase are likely to emit dual jets \cite{Palenzuela:2010nf} that may be detected by future telescopes, such as the Vera C. Rubin Observatory (formerly known as LSST) \cite{LSST:2022kad} and the European Extremely Large Telescope  \cite{Liske:2008ph}. These EM signals can also serve as EM counterparts to provide redshifts \cite{Tamanini:2016zlh}. According to the analysis in Refs.~\cite{Tamanini:2016zlh,Wang:2021srv}, in a 5-year observation, dozens of SMBHBs ($10^4$ -- $10^8$ $M_{\odot}$) with dual-jet EM counterparts could be detected by space-borne observatories in the mHz band. Typically, SMBHBs in the PTA band will inspiral over an extended period, requiring us to wait hundreds of years for the merger phase. Therefore, it is more difficult to detect the merger-phase EM signals for PTA-band SMBHBs.

Since the chirp mass, $M_{\rm c}$, depends on the mass ratio between two black holes forming an SMBHB, the mass ratio affects not only SNRs of GWs but also the probability of the existence of an SMBHB in a galaxy \cite{Chen:2020qlp}. These two effects both affect the constraint precision of cosmological parameters. We define $q=m_1/m_2$ as the mass ratio with $q\in(0,1]$ and set the range of $q$ to [$q_{\rm min}$, 1] according to the log-normal distribution.
The main results of this paper are based on $q_{\rm min}=0.25$ \cite{Mingarelli:2017fbe}. To demonstrate the effect of $q$ more explicitly, we present the constraint results of $H_0$ with various $q_{\rm min}$ values in table~\ref{table3}. We consider four cases, i.e., $q_{\rm min}$ = 1, 0.25, 0.1, and 0.01, where $q_{\rm min}$ = 1 indicates that $q$ is fixed at 1. It is shown that for the bright siren data, the values of $q_{\rm min}$ have negligible effects on constraining $H_0$. Unlike the bright siren data providing almost identical results, the dark siren data constrain $H_0$ more tightly as $q$ decreases. The reason is that when we simulate the dark siren data, $q$ affects not only SNRs of GWs but also the number of mock SMBHBs. The impact of the number of simulated SMBHBs is more significant than the impact of SNRs for the dark siren method.

\begin{table}
\centering
\setlength\tabcolsep{12pt}
\renewcommand\arraystretch{1.5}
\begin{tabular}{ccccccccccc}
\hline\hline
& & \multicolumn{3}{c}{\bf Bright siren}
& & \multicolumn{4}{c}{\bf Dark siren} \\
\cline{3-5} \cline{7-10}
$q_{\rm min}$ & & $N_{\rm s}$
& $\sigma(H_{\rm 0})$ & $\varepsilon(H_{\rm 0})$ & & $N_{\rm mock}$ & $N_{\rm s}$
& $\sigma(H_{\rm 0})$ & $\varepsilon(H_{\rm 0})$ \\ \hline
1     &  & 32 & 0.97 & 0.0142  &   & 54  & 34 & 1.41 & 0.0210  \\
0.25   &  & 25 & 0.98  & 0.0147  &   & 84  & 41 & 1.25 & 0.0184  \\
0.1   &  & 22 & 1.10  & 0.0166  &   & 134 & 65 & 1.21 & 0.0178  \\
0.01 &  & 21 & 1.20  & 0.0171  &   & 197 & 87 & 1.15 & 0.0171  \\
\hline\hline
\end{tabular}
\caption{\label{table3}
  Relative errors of $H_0$ in the $\Lambda$CDM model with different $q_{\rm min}$.
  $N_{\rm mock}$ and $N_{\rm s}$ represent the numbers of mock SMBHBs and detected SMBHBs ($\rho >10$), respectively, and $\varepsilon(H_{\rm 0})$ represents the relative error of $H_0$. Here, we set $N_{\rm p}$ = 100 and $\sigma_t$ = 20 ns.
}
\end{table}

In the analysis of dark sirens, all SMBHBs are simulated at $z<0.05$ based on the 2MASS catalog. The future telescopes, such as the China
Space Station Telescope (CSST) \cite{zhan2021wide}, the Vera C. Rubin Observatory, and the Euclid space mission \cite{EUCLID:2011zbd}, could provide galaxy catalogs at higher redshifts. According to our preliminary estimation, CSST is expected to provide a complete galaxy catalog up to $z\sim 0.3$, where $\mathcal{O}(10^3)$ -- $\mathcal{O}(10^4)$ SMBHBs could be observed by PTAs. Although the measurements on $H_0$ are mainly contributed by the local-Universe SMBHBs considered in this work, a larger number of SMBHB dark sirens may help to measure other cosmological parameters, such as the EoS parameter of dark energy.

Compared with the GW standard sirens in other frequency bands, the ultra-low-frequency GW standard sirens offer some advantages. (i) The masses of the GW sources are at the top of the mass range of SMBHBs, leading to higher SNRs. Figure~\ref{fig4} shows that the highest SNR could reach $\sim$ 700. Such high SNRs are helpful for accurately localizing GW sources, thereby contributing to precise measurements of $H_0$.
(ii) Unlike BNSs that are believed to emit EM signals only during the merger phase, SMBHBs could generate detectable EM signals during the inspiral phase, displaying the characteristic signals of the SMBHB candidates. The inspiral-phase EM signals not only provide redshifts for bright sirens but also serve as early alerts for GW detections, which can assist us in selecting MSPs at optimal sky positions to achieve the highest sensitivity towards the GW source.
(iii) When an SMBHB evolves to the late stage, the GW frequency may fall within the frequency band of space-borne GW detectors. Although most PTA-band SMBHBs inspiral for a long time, in a few cases, for example, an SMBHB with a mass of $ \sim 10^9 M_{\odot}$, $z \lesssim 1$, and $f_0 \sim 10^{-7}$ Hz is expected to enter the merger phase after 17 years. Here, $f_0$ represents the GW frequency at the time of the first observation. Once such cases are discovered, the joint observation in the mHz and nHz frequency bands can be realized. This approach is helpful for localizing GW sources and exploring the diverse physical properties of SMBHBs.

\section{Conclusion}

\label{sec:Conclusion}
The aim of this paper is to explore the potential utilization of ultra-low-frequency GWs as standard sirens. We simulate ultra-low-frequency GWs emitted by individual SMBHBs and analyse the detection capabilities of SKA-era PTAs for these GW signals.
The detected individual SMBHBs can be classified into bright and dark sirens based on whether they have EM counterparts. We simulate the bright siren data using the 154 SMBHB candidates and simulate the dark siren data using the 2MASS catalog. 

We first employ the SMBHB bright siren data to constrain the $\Lambda$CDM model.
In the case of $\sigma_t=100$ ns, as $N_{\rm p}$ increases from 100 to 500, $\varepsilon(H_0)$ decreases from 1.4\% to 1.1\%.
In the case of $N_{\rm p}=100$, as $\sigma_t$ decreases from 100 ns to 20 ns, $\varepsilon(H_0)$ decreases from 1.4\% to 0.98\%. It is shown that reducing $\sigma_t$ is more effective than increasing $N_{\rm p}$ in improving the constraining capability of bright sirens. If $\sigma_t$ could reach 20 ns, 100 MSPs would be sufficient to achieve a measurement precision for $H_0$ comparable to that of the current cosmic distance-ladder observations.
Then we employ the SMBHB bright siren data to constrain the $w$CDM model. Although the PTA data alone cannot constrain $w$ effectively, it can provide tight constraints on $H_0$, thereby breaking the degeneracy between the parameters $w$ and $H_0$. In the case of $N_{\rm p}=100$ and $\sigma_{t}=20$ ns, the combination of the CMB and PTA data gives the relative error $\varepsilon(w)=3.7\%$, which is roughly comparable with the result of \emph{Planck} 2018 TT,TE,EE+lowE+lensing+SNe+BAO. 
Finally, we employ the SMBHB dark siren data to constrain $H_0$ in the $\Lambda$CDM model. It is worth noting that even with only 100 MSPs ($N_{\rm p}$ = 100 and $\sigma_t$ = 20 ns), the precision of $H_0$ could reach $\sim 1.8\%$. Compared with the SMBHB bright sirens, the SMBHB dark sirens have a similar capability of measuring $H_0$. This indicates that even if it is difficult to detect EM counterparts of SMBHBs in the future, dark sirens could solely provide precise measurements of $H_0$. 



We conclude that ultra-low-frequency GWs emitted by individual SMBHBs can serve as both bright and dark sirens, showing promising potential in two aspects. (i) The bright siren data could effectively break the cosmological-parameter degeneracy inherent in the CMB data. The bright siren data, combined with the CMB data, have a comparable capability to the mainstream observational data for measuring $w$.
(ii) The dark sirens in the local Universe have high SNRs and could be well localized, which could result in a measurement precision of $H_0$ that is comparable to that of the current distance-ladder observations. The bright and dark sirens can complement each other to precisely measure both $w$ and $H_0$. Ultra-low-frequency GWs detected by SKA-era PTAs could be developed into a precise late-Universe probe to investigate the nature of dark energy and determine the Hubble constant.

\acknowledgments
We thank Guang-Peng Zhang for his contribution in the early stage of this work.
We are grateful to Zu-Cheng Chen, Ji-Yu Song, and Shang-Jie Jin for fruitful discussions.
This work was supported by the National SKA Program of China (Grants Nos. 2022SKA0110200 and 2022SKA0110203), the National Natural Science Foundation of China (Grants Nos. 12473001, 11975072, 11835009, 11875102, and 12305058), the China Manned Space Program (Grant No. CMS-CSST-2025-A02), the Liaoning Revitalization Talents Program (Grant No. XLYC1905011), the National Program for Support of Top-Notch Young Professionals (Grant No. W02070050), the National 111 Project of China (Grant No. B16009), and the Natural Science Foundation of Hainan Province (Grant No. 424QN215).

\bibliography{main}{}
\bibliographystyle{JHEP}

\end{document}